\documentclass[a4paper,12pt]{article}
\usepackage[USenglish]{babel} 
\usepackage[T1]{fontenc}
\usepackage[ansinew]{inputenc}
\usepackage{lmodern} 
\usepackage{graphicx} 
\usepackage{amsmath}
\usepackage{amsthm}
\usepackage{siunitx} 
\usepackage{amsfonts}
\usepackage{a4wide} 
\usepackage{booktabs}
\usepackage{xspace}
\usepackage{authblk} 
\usepackage{setspace} 
\usepackage{xspace}
\usepackage[noadjust]{cite}
\usepackage{bm} 
\usepackage{color}
\newcommand{\ali}{$\alpha$-Li$_2$IrO$_3$\xspace}
\newcommand{\bli}{$\beta$-Li$_2$IrO$_3$\xspace}
\newcommand{\nairi}{Na$_2$IrO$_3$\xspace}
\newcommand{\liru}{Li$_2$RuO$_3$\xspace}
\newcommand{\lipt}{Li$_2$PtO$_3$\xspace}

\newcommand{\alio}{$\alpha$-Li$_2$IrO$_3$\xspace}

\begin{document}
\title{\vspace{-1.8cm} Single crystal growth from separated educts and its application to lithium transition-metal oxides}
\author[1]{F. Freund}
\author[2]{S. C. Williams}
\author[2]{R. D. Johnson}
\author[2]{R. Coldea}
\author[1]{P. Gegenwart}
\author[1]{A. Jesche}

\singlespacing
\affil[1]{\textit{\small{EP VI, Center for Electronic Correlations and Magnetism, Augsburg University, D-86159 Augsburg, Germany}}}
\affil[2]{\textit{\small{Clarendon Laboratory, University of Oxford, Parks Road, Oxford OX1 3PU, United Kingdom}}}
\renewcommand\Authands{ and }
\date{} 
\maketitle
\onehalfspacing       
\begin{quotation}
\vspace{-.9cm}
\noindent Thorough mixing of the starting materials is the first step of a crystal growth procedure.
This holds true for almost any standard technique, whereas the intentional separation of educts is considered to be restricted to a very limited number of cases.
A noticeable exception is the crystal growth in gels that allows for a better control of the nucleation by limiting the diffusion \cite{Henisch1964}. 
The successful application of this principle to open systems, however, has remained elusive. Here we show that single crystals of \ali can be grown from separated educts in an open crucible in air. Elemental lithium and iridium are oxidized and transported over a distance of typically one centimeter in an isothermal process. Single crystals grow from an exposed condensation point placed in between the educts. The method has also been applied to the growth of Li$_2$RuO$_3$, Li$_2$PtO$_3$ and \bli and a successful use for various other materials is anticipated.
\end{quotation}
The honeycomb iridates \ali and Na$_2$IrO$_3$ attracted a lot of attention after Khaliullin and co-workers proposed that these systems offer a physical realization of the Kitaev interaction \cite{Kitaev2006} in a solid \cite{Jackeli2009, Chaloupka2010}. 
Motivated by their proposal, several experimental studies on single crystalline \nairi and polycrystalline \alio have been performed \cite{Singh2010, Singh2012, Feng2012}.
Direct evidence for the entanglement between spatial and spin directions, which is a consequence of the Kitaev exchange coupling, was recently observed in Na$_2$IrO$_3$ by means of diffuse magnetic X-ray scattering \cite{Chun2015}. These experiments were facilitated by the availability of sizable single crystals \cite{Singh2010}, however, the microscopic details of the growth are not well understood.
\begin{figure*}[t]
	\centering
		\includegraphics[width=0.85\textwidth]{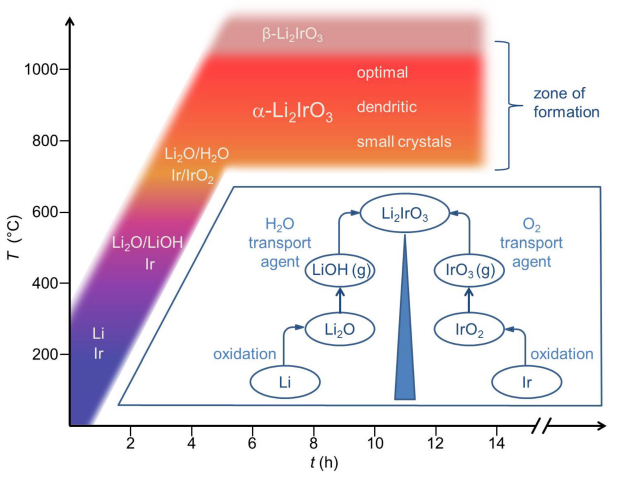}
	\caption{Schematic description of the synthesis method (inset) and temperature profile. Elemental Li and Ir are spatially separated in an open crucible. Upon heating in air Li initially forms solid LiOH that transforms to Li$_2$O at $T >$ \SI{500}{\celsius}. Ir partially oxidizes to IrO$_2$. The formation of $\alpha$-Li$_2$IrO$_3$ takes place at $T = \SI{750}{\celsius} - \SI{1050}{\celsius}$. Single crystals grow from an exposed condensation point placed in between the educts.}
	\label{fig0}
\end{figure*}
For $\alpha$-Li$_2$IrO$_3$ it has not been possible so far to obtain single crystalline material $-$ not even on a length scale of \SI{10}{\micro\metre}. 
Accordingly, there has been no direct access to the anisotropy of the physical properties, and the magnetic structure as well as the contribution of the Kitaev exchange has been still under debate.\\
The growth procedure presented in this letter allows the growth of single crystals of \mbox{$\alpha$-Li$_2$IrO$_3$} of one millimeter along a side. 
A schematic sketch of the synthesis method and the phase formation as function of time and temperature are shown in Fig.\,\ref{fig0}. 
A remarkable feature is the isothermal nature of the process that was revealed by careful temperature measurements at different positions of the crucible (see Supplementary Figure 1): instead of a temperature gradient, here $<1\frac{\text{K}}{\text{cm}}$, it is the formation of $\alpha$-Li$_2$IrO$_3$ itself that drives the transport by maintaining a concentration gradient. The proposed, relevant transport equations are:
\[\hspace{3cm}\text{Li}_2\text{O(s) + H}_2\text{O(g)}\rightleftharpoons 2 \,\text{LiOH(g) \hspace{2cm} \ \cite[p. 166]{Binnewies2012}}\]
\[\hspace{3cm} \ \text{IrO}_2\text{(s)} + \frac{1}{2}\text{O}_2\text{(g)} \rightleftharpoons \text{IrO}_3\text{(g)} \ \hspace{2.5cm} \ \ \text{\cite[p. 217]{Binnewies2012}}\]
Single crystalline \ali forms from gaseous LiOH and IrO$_3$.
The X-ray diffraction pattern and the sharpness of the phase transition to the magnetically ordered state revealed a superior sample quality when compared to polycrystalline material (see below).
Furthermore, the magnetic structure has been solved by recent single crystal magnetic resonant X-ray diffraction measurements performed on these samples (published separately \cite{Steph2016}).
The synthesis of \ali was first reported by Kobayashi \textit{et al.} in 1997\,\cite{Kobayashi97}. 
Polycrystalline material was obtained by heating mixtures of Li$_2$CO$_3$ and IrO$_2$ to temperatures between $\SI{650}{\celsius} - \SI{1050}{\celsius}$.
The presence of a low-spin Ir$^{\,4+}$ state with an effective spin 1/2, as one of the essential ingredients of the Kitaev model, was reported soon after\,\cite{Kobayashi03}.
Despite the substantial interest in this material, the basic synthesis root has not changed since: to the best of our knowledge all attempts made are based on using Li$_2$CO$_3$ as starting material.
Heating mixtures of Li$_2$CO$_3$ with Ir or IrO$_2$ to sufficiently high temperatures leads to the formation of $\alpha$-Li$_2$IrO$_3$ under release of CO$_2$. 
This process, often referred to as 'calcination', has been applied to the growth of several other related materials, e.g.: Li$_2$RuO$_3$ \cite{Dulac1970}, Na$_2$IrO$_3$ \cite{McDaniel1974} or Na$_2$PtO$_3$ \cite{McDaniel1974}.
In this way comparatively large single crystals of Na$_2$IrO$_3$ were obtained \cite{Singh2010,Chun2015}. 
The samples show a plate-like habit with typical lateral dimensions of a few square millimeter and a thickness of 100\,$\mu$m. 
They grow out of a polycrystalline base ('poly bed') and form predominantly at the upper part of the product.  
For $\alpha$-Li$_2$IrO$_3$, however, the similar approach leads to only a fine powder. Different flux methods, especially pre-sintered $\alpha$-Li$_2$IrO$_3$ in LiCl flux, failed to increase the crystal size. 
Nevertheless, a better crystallinity was inferred from X-ray powder diffraction measurements \cite{SohamPhD}. 
In those attempts, the LiCl does not act as a 'classical' flux, it rather promotes a solid state reaction with enhanced diffusion.\\
At temperatures above \SI{1000}{\celsius} the formation of $\alpha$-Li$_2$IrO$_3$ competes with the high-tem\-per\-a\-ture polytype $\beta$-Li$_2$IrO$_3$ \cite{TakayamaBeta}. After repetitive heating of $\alpha$-Li$_2$IrO$_3$ at \SI{1100}{\celsius} small single crystals up to several 10\,\SI{}{\micro\meter} of $\beta$-Li$_2$IrO$_3$ form \cite{BiffinBeta}. Annealing $\beta$-Li$_2$IrO$_3$ at temperatures below \SI{1000}{\celsius} did not lead to the formation of $\alpha$-Li$_2$IrO$_3$, indicating that the transition is irreversible. Small single crystals of a third modification, the 'harmonic' honeycomb $\gamma$-Li$_2$IrO$_3$, were obtained by the calcination of Li$_2$CO$_3$ and IrO$_2$ followed by annealing in molten LiOH at \SIrange{700}{800}{\celsius} \cite{ModicGamma}.
An advantage of the calcination process is the ability to start from carbonates which are comparatively easy to handle and store.  
In contrast, elemental lithium is air sensitive and has been avoided as an educt in previous approaches.
Furthermore, lithium reacts with many standard crucible materials and develops a moderately high vapor pressure (\SI{17}{\milli\bar} at \SI{900}{\celsius} \cite{Honig69}). 
On the other hand, elemental lithium has several advantages for the use as a flux. 
Its low melting point of \SI{180}{\celsius} in comparison with a high boiling temperature of \SI{1342}{\celsius} fulfill two key characteristics of a good flux \cite{Anton}.
Furthermore, lithium has a good solubility for iridium \cite{Massalski1990}.
However, all our attempts to grow single crystals of $\alpha$-Li$_2$IrO$_3$ from a lithium-rich flux and mixtures of lithium with LiCl, LiOH, LiBO$_2$ and/or Li$_2$CO$_3$ failed. A comprehensive overview of those attempts is given in the Supplementary Table 1.\\
Comparatively large single crystals of several millimeter along a side, as observed for Na$_2$IrO$_3$ \cite{Singh2010, Chun2015} are not expected to grow in a solid state reaction due to the limited diffusion length.
Given that the calcination process is completed at these temperatures (at \SI{1050}{\celsius}) and the compound does not melt congruently indicates the relevance of a vapor transport process within the crucible.
In order to investigate the possible formation and transport of Li-O, Ir-O, and/or Li-Ir-O gas specious during the syntheses of $\alpha$-Li$_2$IrO$_3$, we started a growth attempt from elemental lithium and iridium in air.
Lithium granules were placed on iridium powder in an Al$_2$O$_3$ crucible. 
The mixture was heated to \SI{900}{\celsius} over 4\,h, held for 72\,h and quenched to room temperature. 
To our surprise, already the first attempt revealed $\alpha$-Li$_2$IrO$_3$ single crystals of up to \SI{50}{\micro\metre} along a side.
The whole product appeared homogeneous and X-ray powder diffraction pattern showed only small amounts of IrO$_2$ and Ir. 
This is even more surprising since only three small lithium granules (roughly \SI{4}{\milli\metre} in length with a diameter of \SI{1.5} {\milli\metre}) were used but $\alpha$-Li$_2$IrO$_3$ formed over the whole bottom of the crucible (inner diameter of \SI{16}{\milli\metre}).
This observation strongly supports the idea of vapor transport playing a decisive role for the growth of this material. 
However, a classical vapor transport along a temperature gradient does not seem to take place: various growth attempts in a horizontal tube furnace indicated that once $\alpha$-Li$_2$IrO$_3$ has formed it does not transport anymore. This observation is corroborated by an estimate of the free enthalpy of formation for \ali: $\Delta \textit{H}^{0}_{\text{B, 298}} = -880 \frac{\text{kJ}}{\text{mol}} \; \text{and} \; \textit{S}^{0}_{298} = 89 \frac{\text{J}}{\text{mol} \cdot \text{K}}$ [M. Schmidt, MPI-CPfS, private communication]. It corresponds to a large, exothermic value of the free reaction enthalpy of $-332\,\frac{\text{kJ}}{\text{mol}}$ for the following equilibrium equation:
\[\text{IrO}_3\text{(g) + 2 LiOH(g)} \longrightarrow \text{Li}_2\text{IrO}_3\text{(s) + H}_2\text{O(g)} + \frac{1}{2} \text{O}_2\]
Therefore, we started to investigate the growth from spatially separated educts. For this purpose a specially designed setup has been constructed as depicted in Fig.\,\ref{fig1}a$-$c. It consists of a standard crucible, rings (washers), rings with spikes and a disc with a center hole (aperture). All parts are made from Al$_2$O$_3$. The rings act as spacers, hold the aperture in place and allow to vary the distance between starting materials and spikes.\\
The spikes provide an exposed condensation point in between the educts. They are stacked as a 'spiral staircase' in order to identify the ideal position for the growth. The aperture is placed above the spikes and acts as a platform for one of the starting materials. 
The other educt is placed on the bottom of the crucible in the center of two spacers. 
This avoids a direct contact between the material and the spikes which sit on top of the spacers.
\begin{figure*}[t]
	\centering
		\includegraphics[width=.83\linewidth]{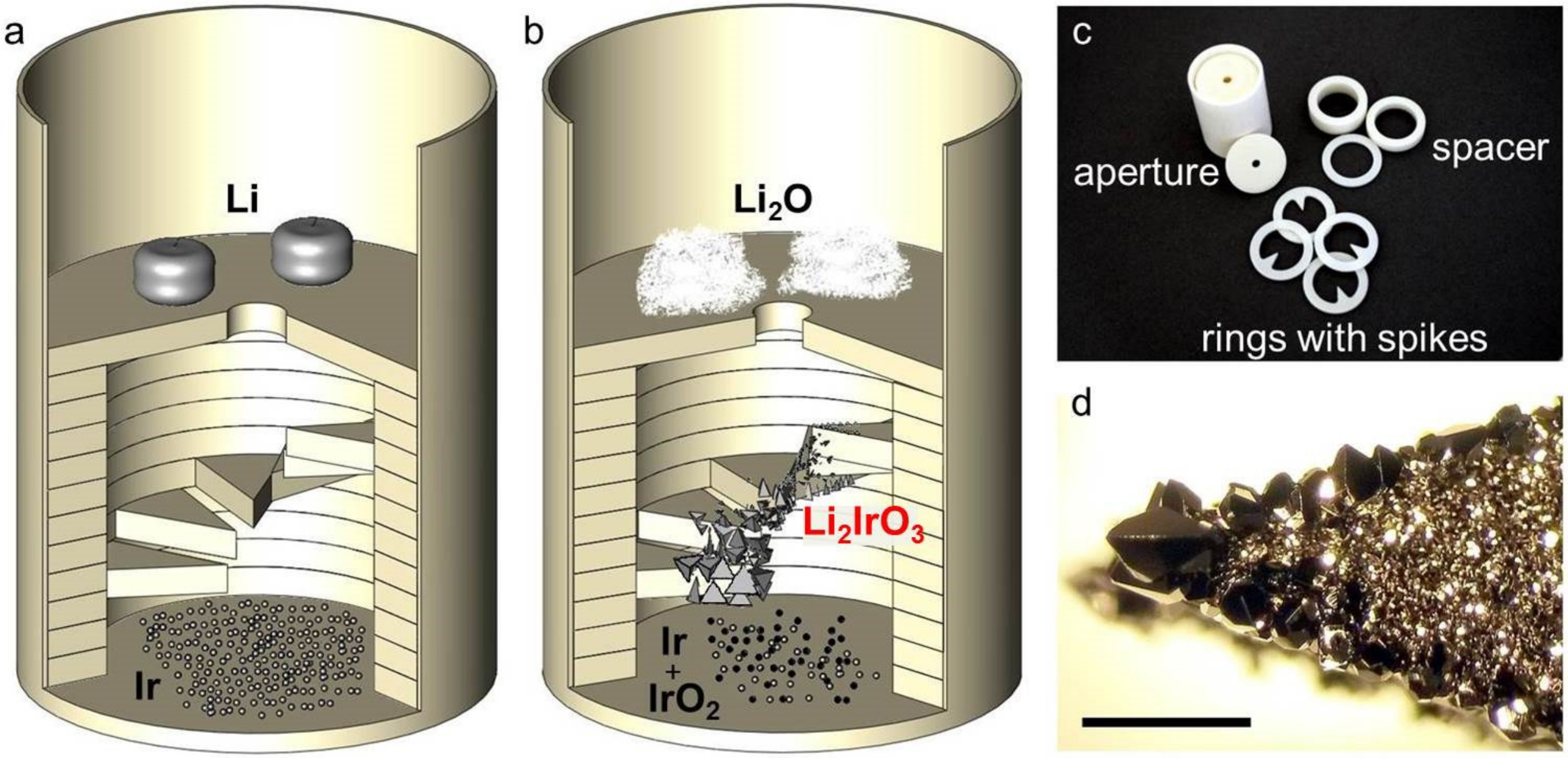}
	\caption{Crystal growth equipment (crucible diameter \SI{16}{\milli\meter}). Arrangement of the materials before and after the growth process is depicted in a) 					and b), respectively. The rings with spikes are oriented like a spiral staircase in order to allow for nucleation at different positions with less
					 intergrowth of the crystals. Formation of the largest $\alpha$-Li$_2$IrO$_3$ single crystals is observed on spikes placed roughly 4\,mm above the Ir 
					 starting material. c) individual setup parts made from Al$_2$O$_3$ and d) typical appearance of one of the lower spikes covered with \ali crystals 
					 at the bottom side, scale bar \SI{1}{\milli\meter}.}
	\label{fig1}
\end{figure*}
\begin{figure}[t]
	\centering
		\includegraphics[width=.83 \linewidth]{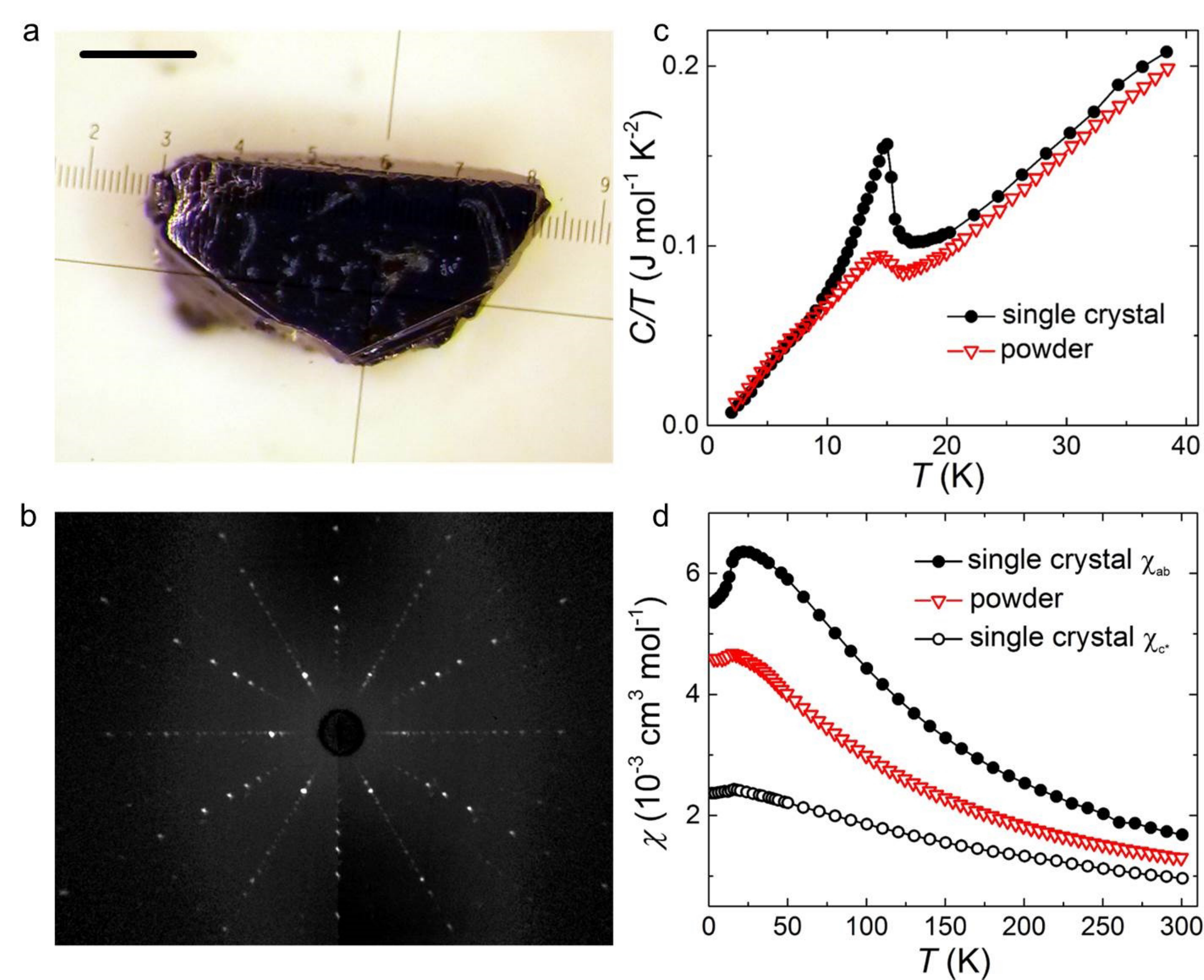}
	\caption{Sample quality and magnetic anisotropy of $\alpha$-Li$_2$IrO$_3$.
					 a) comparatively large single crystal (\SI{1.2 x 0.4 x 0.5}{\milli\metre} and $m$ = \SI{1.7}{\milli\gram}) grown from separated educts (scale bar \SI{0.3}{\milli\metre}).  
					 The corresponding Laue-back-reflection pattern, depicted in b), shows the (nearly) three-fold rotation symmetry perpendicular to the honeycomb layers. c) temperature dependent specific heat of the single crystal shown in a) in comparison with a typical polycrystalline sample. d) an easy-plane anisotropy is apparent from the temperature dependent magnetic susceptibility 
					 ($\mu_0H = 1$\,T, $\chi_{ab}: H \perp c^*$, $\chi_{c^*}: 	H \parallel c^*$).}
	\label{fig2}
\end{figure}
For the growth of $\alpha$-Li$_2$IrO$_3$, iridium metal powder and lithium granules are used as starting materials. Iridium is placed on the bottom of the crucible, lithium on top of the aperture. The distance between the elements is roughly \SI{11}{\milli\meter} with five spikes placed in-between. 
The masses were chosen in their stoichiometric ratio. The whole setup is placed in a box furnace at \SI{200}{\celsius}, heated to \SI{1020}{\celsius} with a rate of \SI{180}{\celsius} per hour, held for three days and finally quenched to room temperature.
While heating, lithium transforms to a Li$_2$O/LiOH mixture at moderate temperatures. At \SI{900}{\celsius} all lithium is burned to Li$_2$O. See a detailed analysis of this process in the Supplementary Figure 2. Only small amounts of Li$_2$O are found on top of the aperture (where the lithium was placed) after the process. The iridium powder placed at the bottom is partially oxidized to IrO$_2$.
$\alpha$-Li$_2$IrO$_3$ covers large parts of the spikes, with the largest crystals growing at the tip of the spikes, $3 - 4$\,mm above the iridium (Fig.\,\ref{fig1}d).\\
Single crystals of dimensions larger than 1\,mm were obtained (Fig.\,\ref{fig2}a).
A good sample quality is inferred from Laue-back-reflection (Fig.\,\ref{fig2}b): the diffraction pattern shows the (nearly) three-fold rotational symmetry of the honeycomb layers (along the $\bm{c}^*$-direction). The spot-size of the X-ray beam was similar to the sample dimensions. 
Figure\,\ref{fig2}c shows the temperature dependent specific heat measured on the same single crystal in comparison with polycrystalline material that was grown by calcination \cite{Singh2012}. The improved sample quality of the single crystal is apparent from a sharper transition to the antiferromagnetically ordered state at $T_{\rm N} = 15$\,K. Temperature-dependent magnetic susceptibility for field applied parallel ($\chi_{ab}$) and perpendicular ($\chi_{c^*}$) to the honeycomb layers is shown in Fig.\,\ref{fig2}d. An easy-plane behavior with a sharp decrease of $\chi_{ab}$ at $T_{\rm N}$ is observed, whereas $\chi_{c^*}$ decreases only slightly. 
A measurement performed on polycrystalline material \cite{Singh2012} is included for comparison and can be roughly described by $1/3\,\chi_{c^*} + 2/3\,\chi_{ab}$ for $T > T_{\rm N}$.
\begin{figure*}[t]
\centering
\includegraphics[width=0.93\textwidth]{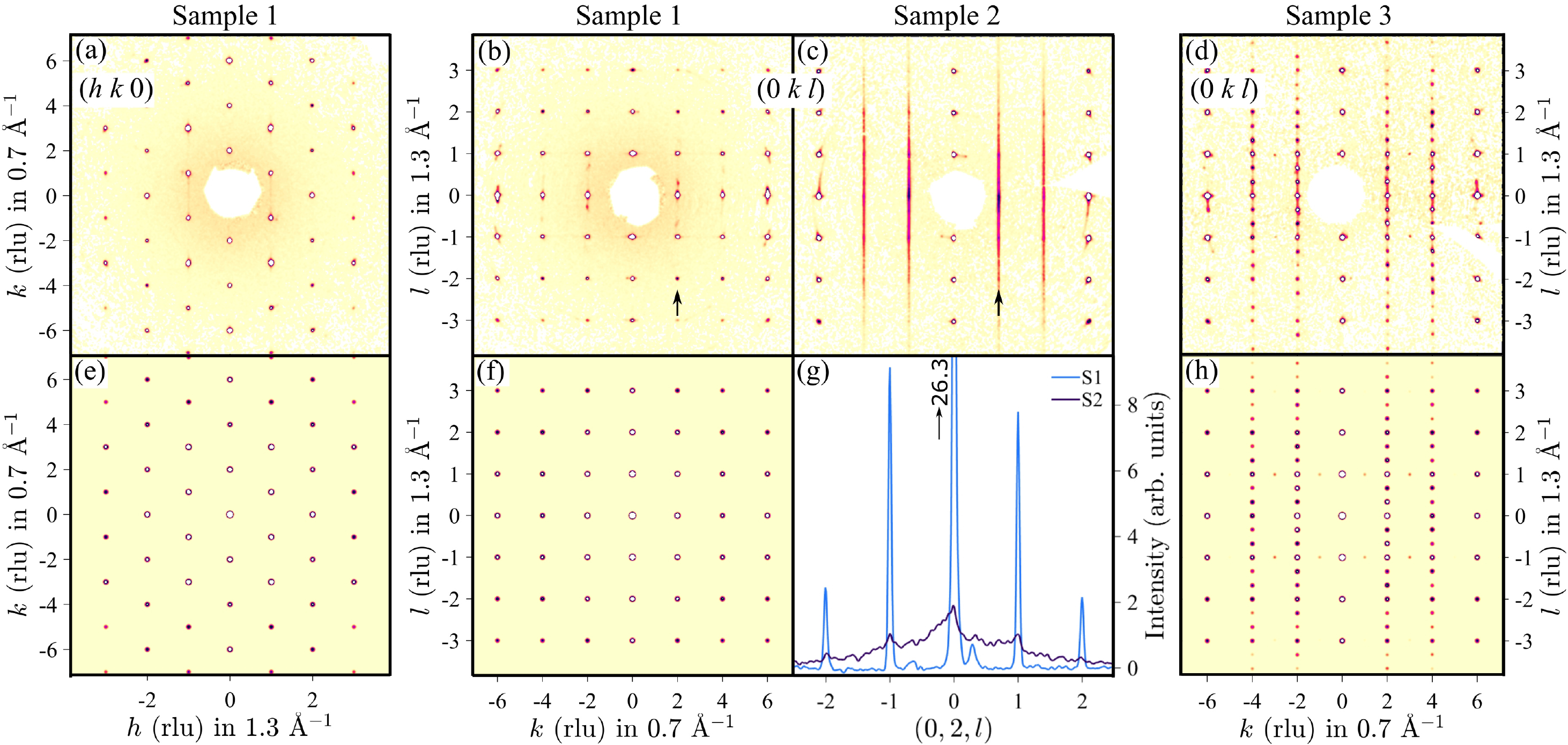}
\caption[]{
X-ray diffraction pattern from three different \alio crystals: one un-twinned
and without stacking faults (sample 1, panels a$-$b), one predominantly a single grain,
but with significant stacking faults manifested in extended diffuse scattering
along \textit{l} (sample 2, panel c), and one multi-twin crystal (sample 3, panel d).
Color is intensity on a log scale. Vertical arrows near $k=2$ in panels b$-$c show direction
along which the intensity is plotted in panel g, note the strong
contrast between sample 1 with sharp peaks at integer $l$ and
sample 2 where diffuse scattering dominates. Bottom graphs e, f, h
show the calculated X-ray diffraction pattern in the same axes as
the above panels a, b, d, for the nominal monoclinic crystal
structure of \alio \cite{OMalley2008}. Panel h includes
contribution from C$^{\pm}$ twins (grains rotated by
$\pm120^{\circ}$ around $\bm{c}^*$ leading to the peaks at
fractional coordinates ($0,k,n\pm1/3$), with $n$ integer and
$k=\pm2,\pm4$) and an A-type twin responsible for the peaks at
($0,k,\pm1$) with $k$ odd (see Supplementary Note for
details).} 
\label{fig:xrays}
\end{figure*}
\\
The structural order in the grown crystals was probed using X-ray
diffraction and representative patterns are shown in Fig.\,\ref{fig:xrays}a$-$d. The
data are fully consistent with the expected monoclinic crystal
structure \cite{OMalley2008} of alternate stacking of honeycomb
Li$_{1/2}$IrO$_3$ and hexagonal Li layers with space group $C2/m$
(calculated patterns shown in Fig.\,\ref{fig:xrays}e$-$f). The unit
cell parameters obtained from indexing the diffraction data are
$a=5.169(7)$\AA{}, $b=8.938(8)$\AA{}, $c=5.121(7)$\AA{},
$\beta=109.75(6)^{\circ}$, consistent with the earlier powder data
\cite{OMalley2008}. Samples grown at \SI{900}{\celsius} showed pronounced rods of diffuse scattering along the ${\bm c}^*$
direction, normal to the layers, as evidenced in
Fig.\,\ref{fig:xrays}c. Rods of diffuse scattering with the same
selection rule were also observed in the iso-structural materials
Na$_2$IrO$_3$ and $\alpha$-RuCl$_3$ \cite{Johnson2015} and
attributed \cite{Choi2012} to occasional in-plane shifts of the
stacked layers by $\pm{\bm{b}}/3$. Monitoring the structural order
for different growth temperatures allowed us to optimize the
synthesis parameters and obtain crystals with almost no detectable
diffuse scattering (compare Fig.\,\ref{fig:xrays}b$-$c, and the
intensity profile in Fig.\,\ref{fig:xrays}g), showing that those
crystals grown at \SI{1020}{\celsius} are close to the limit of fully-coherent,
three-dimensional structural ordering. Fig.\,\ref{fig:xrays}a$-$b
show data from an un-twinned single crystal. Most as-grown
crystals are twinned and a representative diffraction pattern
shown in Fig.~\ref{fig:xrays}d can be understood by three
additional twins: two twins rotated by $\pm120^{\circ}$ around
$\bm{c}^*$, and another twin with the $\bm{a}$ and $\bm{c}$ axes
interchanged (for more details see Supplementary Note). We
note that the susceptibility data in Fig.\,\ref{fig2}d was collected on a
crystal that contained predominantly twins rotated by
$\pm120^{\circ}$ around ${\bm c}^*$, so under the assumption that
the susceptibility tensor has only one unique axis ${\bm c}^*$
(normal to the $ab$ plane), all those twins had the same magnetic
response in field applied along ${\bm c}^*$ or perpendicular.
\begin{figure}[tbp]
	\centering
		\includegraphics[width=\linewidth]{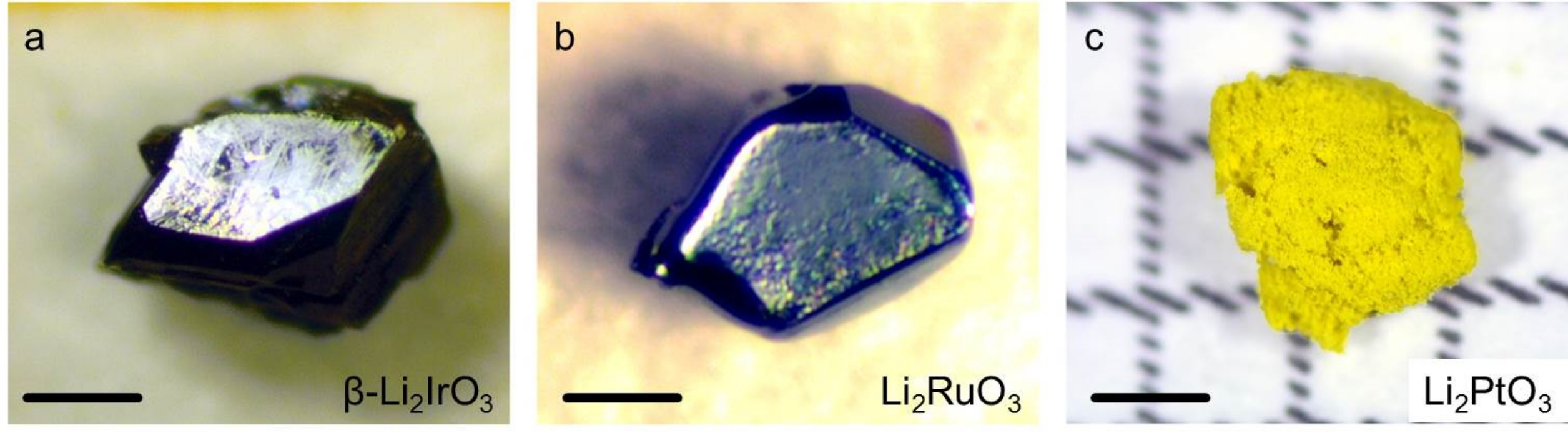}
		\caption{Further materials synthesized from separated educts.
						 a) $\beta$-Li$_2$IrO$_3$ single crystal (scale bar \SI{0.2}{\milli\metre}) and b) a single
						 crystal of Li$_2$RuO$_3$ (scale bar \SI{0.1}{\milli\metre}). 
						 For Li$_2$PtO$_3$ fine yellow powder could be obtained shown in c) (scale bar \SI{1}{\milli\metre}).}
	\label{fig4}
\end{figure}
\\
The method described is not restricted to the growth of $\alpha$-Li$_2$IrO$_3$. 
Single crystals of $\beta$-Li$_2$IrO$_3$ and Li$_2$RuO$_3$ were also obtained (Fig.\,\ref{fig4}a, b, see Supplementary Figure 3 for X-ray diffraction pattern). Formation of the latter is expected from the similar transport behavior of Ir and Ru \cite[p. 214 ff]{Binnewies2012}. For Li$_2$PtO$_3$ we obtained polycrystalline material of good quality (Fig.\,\ref{fig4}c, see Supplementary Figure 3 for X-ray diffraction pattern).  
In conclusion, the technique should be applicable to various transport active elements in air in its simplest form. 
Application to a broader class of materials could be achieved by providing a controlled atmosphere (static or flowing) of, for example, oxygen, chlorine or iodine. 
The combination of an isothermal vapor transport in an open crucible with separated educts is unique and provides another approach for the crystal growth community.\\
\newpage
\noindent 
The authors thank M. Schmidt, V. Tsurkan, A. Tsirlin, S. Manni, G. Hammerl, A. Erb, H. S. Jeevan, C. Krellner, S. Wurmehl, C. Geibel, T. Wolf, D. Schmitz and W. Scherer for fruitful discussions. For the technical support we like to thank A. Mohs, A. Herrnberger and K. Wiedenmann. This work has been supported by the German Science Foundation through projects TRR-80, SPP1666 and the Emmy Noether program - JE 748/1. Work in Oxford was partially supported by the EPSRC (UK) under Grants No. EP/H014934/1 and EP/M020517/1. R.D.J. acknowledges support from a Royal Society University Research Fellowship. In accordance with the EPSRC policy framework on research data, access to the data generated using EPSRC funds will be made available from X-ray data depository at http://dx.doi.org/10.5287/bodleian:O5YMYxv8R.\\
\\
email: anton.jesche@physik.uni-augsburg.de
\singlespacing

\begin{thebibliography}{10}
\expandafter\ifx\csname url\endcsname\relax
  \def\url#1{\texttt{#1}}\fi
\expandafter\ifx\csname urlprefix\endcsname\relax\def\urlprefix{URL }\fi
\providecommand{\bibinfo}[2]{#2}
\providecommand{\eprint}[2][]{\url{#2}}

\bibitem{Henisch1964}
\bibinfo{author}{Henisch, H.~K.}, \bibinfo{author}{Dennis, J.} \&
  \bibinfo{author}{Hanoka, J.~I.}
\newblock \bibinfo{title}{Crystal growth in gels}.
\newblock \emph{\bibinfo{journal}{J. Phys. Chem. Solids}}
  \textbf{\bibinfo{volume}{26}}, \bibinfo{pages}{493--500}
  (\bibinfo{year}{1964}).

\bibitem{Kitaev2006}
\bibinfo{author}{Kitaev, A.}
\newblock \bibinfo{title}{Anyons in an exactly solved model and beyond}.
\newblock \emph{\bibinfo{journal}{Ann. Phys.}} \textbf{\bibinfo{volume}{321}},
  \bibinfo{pages}{2--111} (\bibinfo{year}{2006}).

\bibitem{Jackeli2009}
\bibinfo{author}{Jackeli, G.} \& \bibinfo{author}{Khaliullin, G.}
\newblock \bibinfo{title}{Mott insulators in the strong spin-orbit coupling
  limit: {from} {Heisenberg} to a quantum compass and {Kitaev} models}.
\newblock \emph{\bibinfo{journal}{Phys. Rev. Lett.}}
  \textbf{\bibinfo{volume}{102}}, \bibinfo{pages}{017205}
  (\bibinfo{year}{2009}).

\bibitem{Chaloupka2010}
\bibinfo{author}{Chaloupka, J.}, \bibinfo{author}{Jackeli, G.} \&
  \bibinfo{author}{Khaliullin, G.}
\newblock \bibinfo{title}{{Kitaev-Heisenberg} model on a honeycomb lattice:
  {possible} exotic phases in iridium oxides {${A}_{2}{\mathrm{IrO}}_{3}$}}.
\newblock \emph{\bibinfo{journal}{Phys. Rev. Lett.}}
  \textbf{\bibinfo{volume}{105}}, \bibinfo{pages}{027204}
  (\bibinfo{year}{2010}).

\bibitem{Singh2010}
\bibinfo{author}{Singh, Y.} \& \bibinfo{author}{Gegenwart, P.}
\newblock \bibinfo{title}{Antiferromagnetic {Mott} insulating state in single
  crystals of the honeycomb lattice material
  {${\text{Na}}_{2}{\text{IrO}}_{3}$}}.
\newblock \emph{\bibinfo{journal}{Phys. Rev. B}} \textbf{\bibinfo{volume}{82}},
  \bibinfo{pages}{064412} (\bibinfo{year}{2010}).

\bibitem{Singh2012}
\bibinfo{author}{Singh, Y.} \emph{et~al.}
\newblock \bibinfo{title}{Relevance of the {Heisenberg-Kitaev} model for the
  honeycomb lattice iridates {${A}_{2}{\mathrm{IrO}}_{3}$}}.
\newblock \emph{\bibinfo{journal}{Phys. Rev. Lett.}}
  \textbf{\bibinfo{volume}{108}}, \bibinfo{pages}{127203}
  (\bibinfo{year}{2012}).

\bibitem{Feng2012}
\bibinfo{author}{Ye, F.} \emph{et~al.}
\newblock \bibinfo{title}{Direct evidence of a zigzag spin-chain structure in
  the honeycomb lattice: a neutron and x-ray diffraction investigation of
  single-crystal {Na${}_{2}$IrO${}_{3}$}}.
\newblock \emph{\bibinfo{journal}{Phys. Rev. B}} \textbf{\bibinfo{volume}{85}},
  \bibinfo{pages}{180403} (\bibinfo{year}{2012}).

\bibitem{Chun2015}
\bibinfo{author}{Chun, S.~H.} \emph{et~al.}
\newblock \bibinfo{title}{Direct evidence for dominant bond-directional
  interactions in a honeycomb lattice iridate \nairi}.
\newblock \emph{\bibinfo{journal}{Nat. Phys.}} \textbf{\bibinfo{volume}{11}},
  \bibinfo{pages}{462--466} (\bibinfo{year}{2015}).

\bibitem{Binnewies2012}
\bibinfo{author}{Binnewies, M.}, \bibinfo{author}{Glaum, R.},
  \bibinfo{author}{Schmidt, M.} \& \bibinfo{author}{Schmidt, P.}
\newblock \emph{\bibinfo{title}{{Chemical Vapor Transport Reactions}}}
  (\bibinfo{publisher}{de Gruyter, Berlin/Boston}, \bibinfo{year}{2012}).

\bibitem{Steph2016}
\bibinfo{author}{Williams, S.~C.} \emph{et~al.}
\newblock \bibinfo{title}{Incommensurate counterrotating magnetic order
  stabilized by {Kitaev} interactions in the layered honeycomb {\ali}}.
\newblock \emph{\bibinfo{journal}{Phys. Rev. B}} \textbf{\bibinfo{volume}{93}},
  \bibinfo{pages}{195158} (\bibinfo{year}{2016}).


\bibitem{Kobayashi97}
\bibinfo{author}{Kobayashi, H.} \emph{et~al.}
\newblock \bibinfo{title}{Structure and charge/discharge characteristics of new
  layered oxides: {Li$_{1.8}$Ru$_{0.6}$Fe$_{0.6O3}$ and Li$_2$IrO$_3$}}.
\newblock \emph{\bibinfo{journal}{J. Power Sources}}
  \textbf{\bibinfo{volume}{68}}, \bibinfo{pages}{686--691}
  (\bibinfo{year}{1997}).

\bibitem{Kobayashi03}
\bibinfo{author}{Kobayashi, H.}, \bibinfo{author}{Tabuchi, M.},
  \bibinfo{author}{Shikano, M.}, \bibinfo{author}{Kageyama, H.} \&
  \bibinfo{author}{Kanno, R.}
\newblock \bibinfo{title}{Structure{,} and magnetic and electrochemical
  properties of layered oxides{,} {Li$_2$IrO$_3$}}.
\newblock \emph{\bibinfo{journal}{J. Mater. Chem.}}
  \textbf{\bibinfo{volume}{13}}, \bibinfo{pages}{957--962}
  (\bibinfo{year}{2003}).

\bibitem{Dulac1970}
\bibinfo{author}{Dulac, J.~F.}
\newblock \bibinfo{title}{Synthesis and crystallographic structure of a new
  ternary compound {Li$_2$RuO$_3$}}.
\newblock \emph{\bibinfo{journal}{C. R. l'Acad. Sci., Ser. B}}
  \textbf{\bibinfo{volume}{270}}, \bibinfo{pages}{223--226}
  (\bibinfo{year}{1970}).

\bibitem{McDaniel1974}
\bibinfo{author}{McDaniel, C.~L.}
\newblock \bibinfo{title}{Phase relations in the systems {Na$_2$O-IrO$_2$ and
  Na$_2$O-PtO$_2$ in air}}.
\newblock \emph{\bibinfo{journal}{J. Solid State Chem.}}
  \textbf{\bibinfo{volume}{9}}, \bibinfo{pages}{139--146}
  (\bibinfo{year}{1974}).

\bibitem{SohamPhD}
\bibinfo{author}{Manni, S.}
\newblock \emph{\bibinfo{title}{{Synthesis and investigation of frustrated
  Honeycomb lattice iridates and rhodates}}}.
\newblock Ph.D. thesis, \bibinfo{school}{Georg-August-Universit\"{a}t
  G\"{o}ttingen} (\bibinfo{year}{2014}).

\bibitem{TakayamaBeta}
\bibinfo{author}{Takayama, T.} \emph{et~al.}
\newblock \bibinfo{title}{Hyperhoneycomb iridate \bli as a platform for
  {Kitaev} magnetism}.
\newblock \emph{\bibinfo{journal}{Phys. Rev. Lett.}}
  \textbf{\bibinfo{volume}{114}}, \bibinfo{pages}{077202}
  (\bibinfo{year}{2015}).

\bibitem{BiffinBeta}
\bibinfo{author}{Biffin, A.} \emph{et~al.}
\newblock \bibinfo{title}{Unconventional magnetic order on the hyperhoneycomb
  {Kitaev} lattice in \bli: {full} solution via magnetic resonant x-ray
  diffraction}.
\newblock \emph{\bibinfo{journal}{Phys. Rev. B}} \textbf{\bibinfo{volume}{90}},
  \bibinfo{pages}{205116} (\bibinfo{year}{2014}).

\bibitem{ModicGamma}
\bibinfo{author}{Modic, K.~A.} \emph{et~al.}
\newblock \bibinfo{title}{Realization of a three-dimensional spin--anisotropic
  harmonic honeycomb iridate}.
\newblock \emph{\bibinfo{journal}{Nat. Commun.}} \textbf{\bibinfo{volume}{5}}
  (\bibinfo{year}{2014}).

\bibitem{Honig69}
\bibinfo{author}{Honig, R.~E.} \& \bibinfo{author}{Kramer, D.~A.}
\newblock \bibinfo{title}{Vapor pressure data for solid and liquid elements}.
\newblock \emph{\bibinfo{journal}{RCA Rev.}} \textbf{\bibinfo{volume}{30}},
  \bibinfo{pages}{285} (\bibinfo{year}{1969}).

\bibitem{Anton}
\bibinfo{author}{Jesche, A.} \& \bibinfo{author}{Canfield, P.~C.}
\newblock \bibinfo{title}{Single crystal growth from light, volatile and
  reactive materials using lithium and calcium flux}.
\newblock \emph{\bibinfo{journal}{Phil. Mag.}} \textbf{\bibinfo{volume}{94}},
  \bibinfo{pages}{2372--2402} (\bibinfo{year}{2014}).

\bibitem{Massalski1990}
\bibinfo{author}{Massalski, T.~B.}, \bibinfo{author}{Okamoto, H.} \&
  \bibinfo{author}{Subramanian, P.~R.}
\newblock \emph{\bibinfo{title}{{Binary Alloy Phase Diagrams, 2nd Edition
  (Volume 3)}}} (\bibinfo{publisher}{A. S. M. International, Materials Park,
  OH}, \bibinfo{year}{1990}).

\bibitem{OMalley2008}
\bibinfo{author}{O'Malley, M.~J.}, \bibinfo{author}{Verweij, H.} \&
  \bibinfo{author}{Woodward, P.~M.}
\newblock \bibinfo{title}{Structure and properties of ordered {Li$_2$IrO$_3$
  and Li$_2$PtO$_3$}}.
\newblock \emph{\bibinfo{journal}{J. Solid State Chem.}}
  \textbf{\bibinfo{volume}{181}}, \bibinfo{pages}{1803--1809}
  (\bibinfo{year}{2008}).

\bibitem{Johnson2015}
\bibinfo{author}{Johnson, R.~D.} \emph{et~al.}
\newblock \bibinfo{title}{{Monoclinic crystal structure of $\alpha$-RuCl$_3$
  and the zigzag antiferromagnetic ground state}}.
\newblock \emph{\bibinfo{journal}{Phys. Rev. B}} \textbf{\bibinfo{volume}{92}},
  \bibinfo{pages}{235119} (\bibinfo{year}{2015}).

\bibitem{Choi2012}
\bibinfo{author}{Choi, S.~K.} \emph{et~al.}
\newblock \bibinfo{title}{Spin waves and revised crystal structure of honeycomb
  iridate \nairi}.
\newblock \emph{\bibinfo{journal}{Phys. Rev. Lett.}}
  \textbf{\bibinfo{volume}{108}}, \bibinfo{pages}{127204}
  (\bibinfo{year}{2012}).

\end{thebibliography}

\vspace{-0.2cm}
\begin{thebibliography}{10}
\vspace{-0.3cm}
\bibitem{O'Malley08}
O'Malley, M. J., Verweij, H. \& Woodward, P. M. Structure and properties of ordered
Li$_2$IrO$_3$ and \lipt. \textit{J. Solid State Chem}. \textbf{181}, $1803-1809$ (2008)
\bibitem{Choi12}
Choi, S. K. \textit{et al.} Spin waves and revised crystal structure of honeycomb iridate \nairi. 
\textit{Phys. Rev. Lett.} \textbf{108}, 127204 (2012).
\end{thebibliography}

\setcounter{figure}{0}
\setcounter{table}{0}
\renewcommand{\figurename}{Supplementary Figure}
\renewcommand{\tablename}{Supplementary Table}
\begin{table*}[tph]
\centering
\Large{Supplementary Information to ''Single crystal growth from separated educts and its application to lithium transition-metal oxides''}
\singlespacing
\normalsize
\vspace{0.5cm}
\caption{Results of various attempts to grow \ali single crystals from flux.}
\label{tabflux}
\begin{tabular}{p{0.08\textwidth}p{0.13\textwidth}p{0.1\textwidth}p{0.22\textwidth}p{0.34\textwidth} }
 Flux								&	Starting \ \ \ \ Material		& Crucible						&	Temperature \hspace{1cm} Profile			        		& Result												\\
\toprule			
&\\

Li 									& \ali		& Ta									& 4\,h to \SI{1000}{\celsius}																						& crucible decomposed						\\
&\\
Li 									& \ali		& Nb									& 4\,h to \SI{1000}{\celsius}																						& crucible decomposed						\\
&\\
Li			 						& IrO$_2$	&	Nb									& 5\,h to \SI{1000}{\celsius}; hold 5\,h; 60\,h to \SI{200}{\celsius}		& Ir single crystals, 
																																																															crucible \mbox{attacked} 			\\
&\\
Li, LiBO$_2$ 				& \ali		& Pt \hspace{1cm} in quartz		& 5\,h to \SI{950}{\celsius}; hold 1\,h; 60\,h to \SI{500}{\celsius}		& quartz tube broken,
																																																															\ali transformed to IrO$_2$		\\
&\\
LiCl, Li$_2$CO$_3$	&	\ali 		& Al$_2$O$_3$  \ \  in quartz & 2\,h to \SI{1000}{\celsius}; hold 2\,h; 55\,h to \SI{500}{\celsius}		& well ordered \ali powder, \bli
																																																															impurities 										\\
&\\ 
LiCl, LiOH					&	\ali 		& Al$_2$O$_3$ \ \ in quartz & 2\,h to \SI{1000}{\celsius}; hold 2\,h; 55\,h to \SI{500}{\celsius}		& well ordered \ali, \mbox{\bli}																																																																		and IrO$_2$						   \\
&\\
LiCl, Li$_2$CO$_3$	&	\ali 		& Al$_2$O$_3$ \ \ in quartz & 2\,h to \SI{850}{\celsius}; hold 2\,h; 55\,h to \SI{500}{\celsius}		& \ali pellet not solved in flux\\
&\\ 
LiCl, LiOH					&	\ali 		& Al$_2$O$_3$ \ \ in quartz & 2\,h to \SI{850}{\celsius}; hold 2\,h; 55\,h to \SI{500}{\celsius}		& \ali powder 									\\
&\\
LiOH								&	\ali 		& Al$_2$O$_3$ \ \ in Nb 		& 8\,h to \SI{800}{\celsius}; hold 6\,h; 50\,h to \SI{500}{\celsius} 		& \mbox{\ali powder,}
																																																																		\mbox{Ir impurities}		\\
&\\
LiCl								&	\ali 		& Al$_2$O$_3$ \ \ in quartz & 2\,h to \SI{850}{\celsius}; hold 2\,h; 55\,h to \SI{500}{\celsius}		& \ali powder, \bli and IrO$_2$
																																																															impurities										\\
&\\
\bottomrule
\end{tabular}
\end{table*}
\newpage
\begin{figure}[tp]
	\centering
		\includegraphics[width=\textwidth]{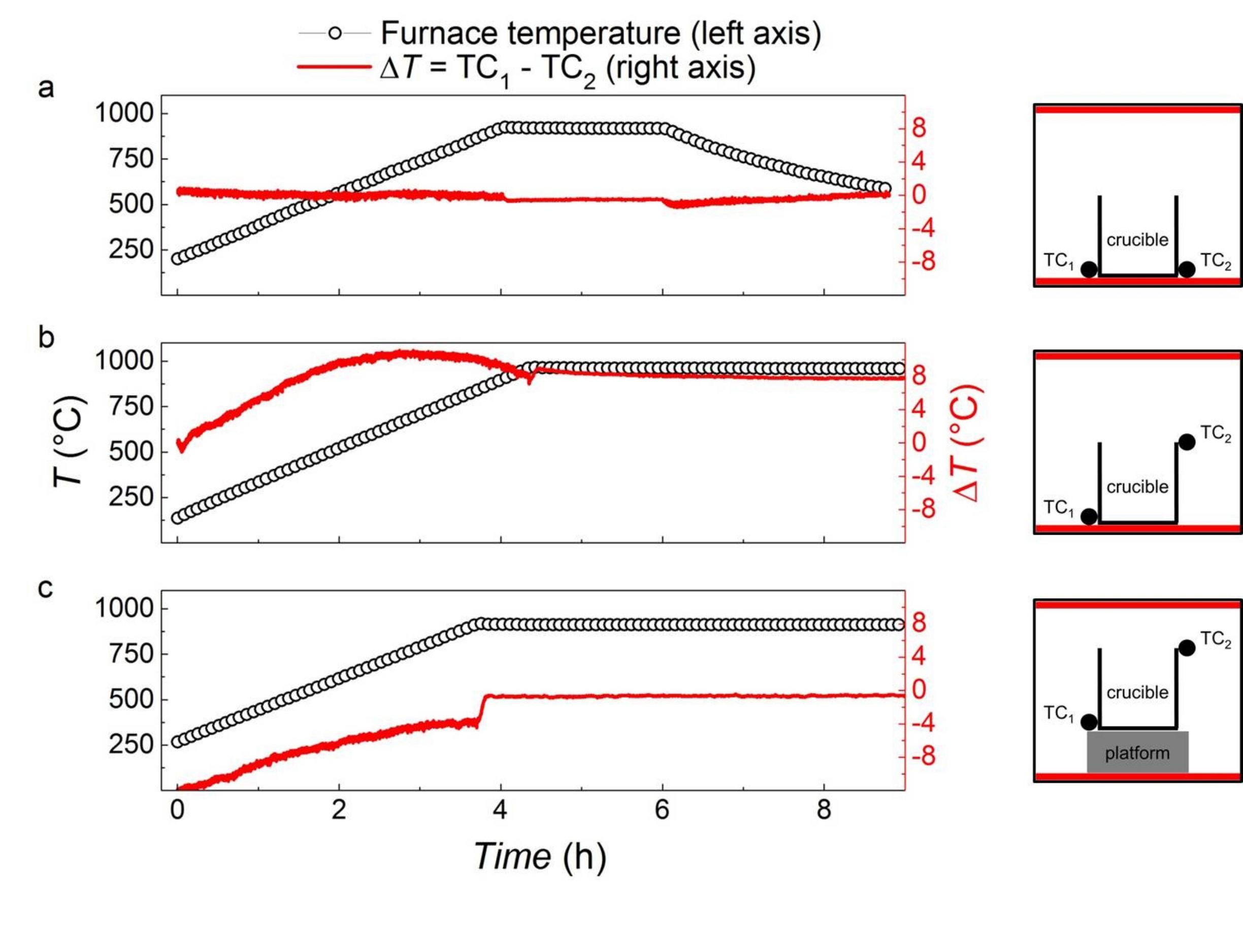}
	\caption{Temperature gradient between bottom and top of the crucible during the growth. Temperature profiles are shown on the left hand side. The position of the thermocouples (TCs) are depicted on the right. The heating plates of the furnace are placed on the top and bottom (marked in red). a), verification of the calibration of the used thermocouples. TC$_1$ and TC$_2$ are both placed on the bottom of the crucible. The crucible stands on the lower heating plate. A small difference in the range of $\Delta T$ $\approx$ $ \pm \, \SI{1}{\celsius}$ is confirmed.
	b), TC$_1$ stays at the bottom and TC$_2$ is placed on the top. The gradient increases during the heating process. At the final temperature the difference is constant and small with $\frac{\Delta \textit{T}}{\Delta \textit{z}} \approx \frac{8\,\text{K}}{2\,\text{cm}} = 4 \frac{\text{K}}{\text{cm}}$. 
	c), in order to investigate the effect of the small temperature gradient observed in b), the whole setup is placed on a platform to equalize the distance of the crucible to the upper and lower heating plate of the furnace. As expected for a symmetric arrangement $\Delta T \approx$ \SI{0}{\kelvin}. The growth is not affected and it is still possible to grow \ali single crystals in this configuration. This shows the minor importance of a temperature gradient in this method.}
	\label{fig:sub1}
\end{figure}

\begin{figure}[tp]
	\centering
		\includegraphics[width=0.8\textwidth]{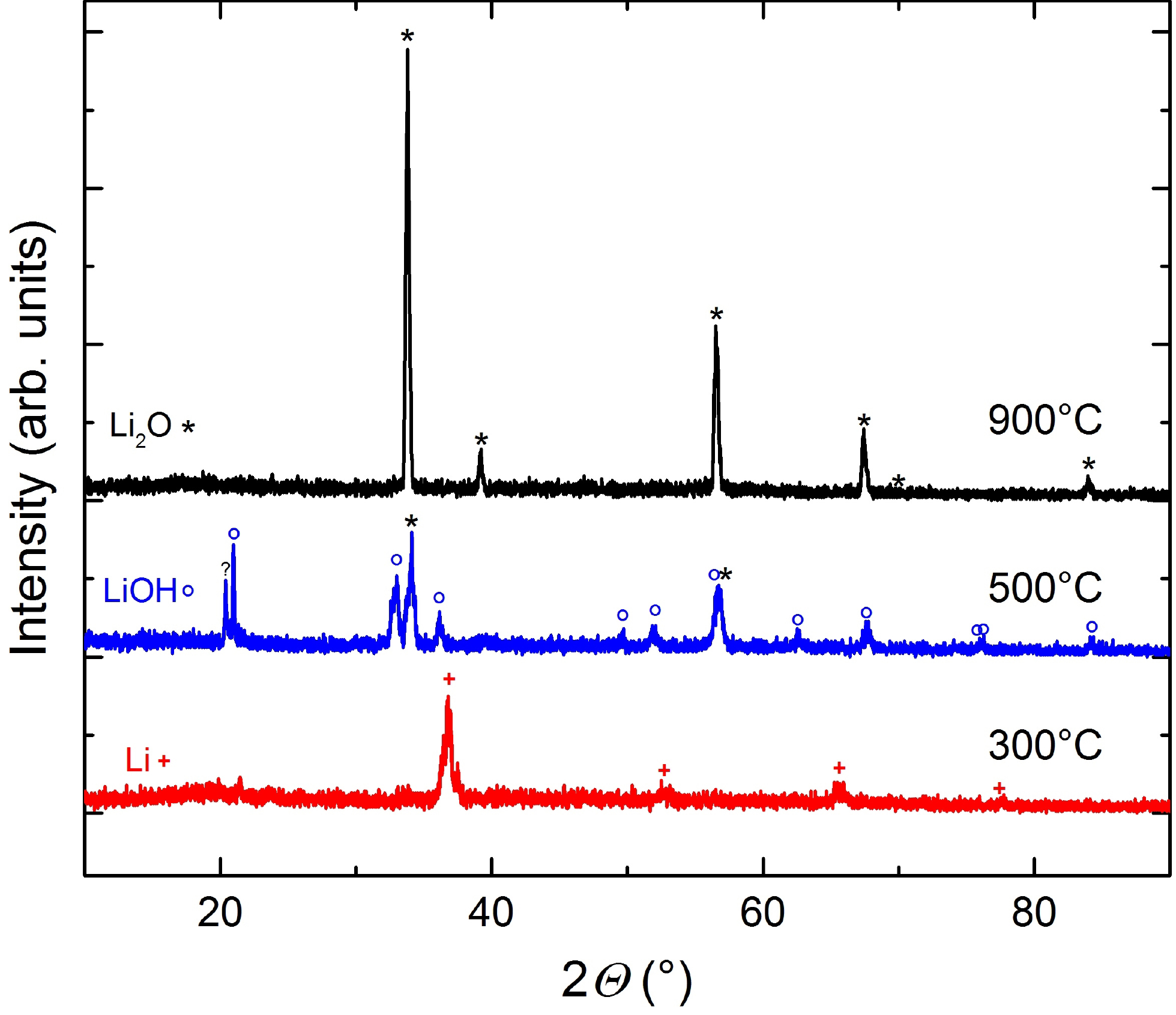}
	\caption{Evolution of the lithium educt during the heating process. Three pieces of elemental lithium are placed in three alumina crucibles, put in an open box furnace at \SI{200}{\celsius} and heated to \SI{900}{\celsius} in four hours. The samples are taken out of the furnace during heating at \SI{300}{\celsius}, \SI{500}{\celsius} and \SI{900}{\celsius}, respectively. Powder X-ray diffraction patterns, collected immediately after taking out the sample, show the transformation of lithium to a Li$_2$O-LiOH mix, ending in pure Li$_2$O. The theoretical peak positions of Li, Li$_2$O and LiOH are indicated in the patterns.}
	\label{fig:sub3}
\end{figure}

\begin{figure}[tbp]
	\centering
		\includegraphics[width=0.8\textwidth]{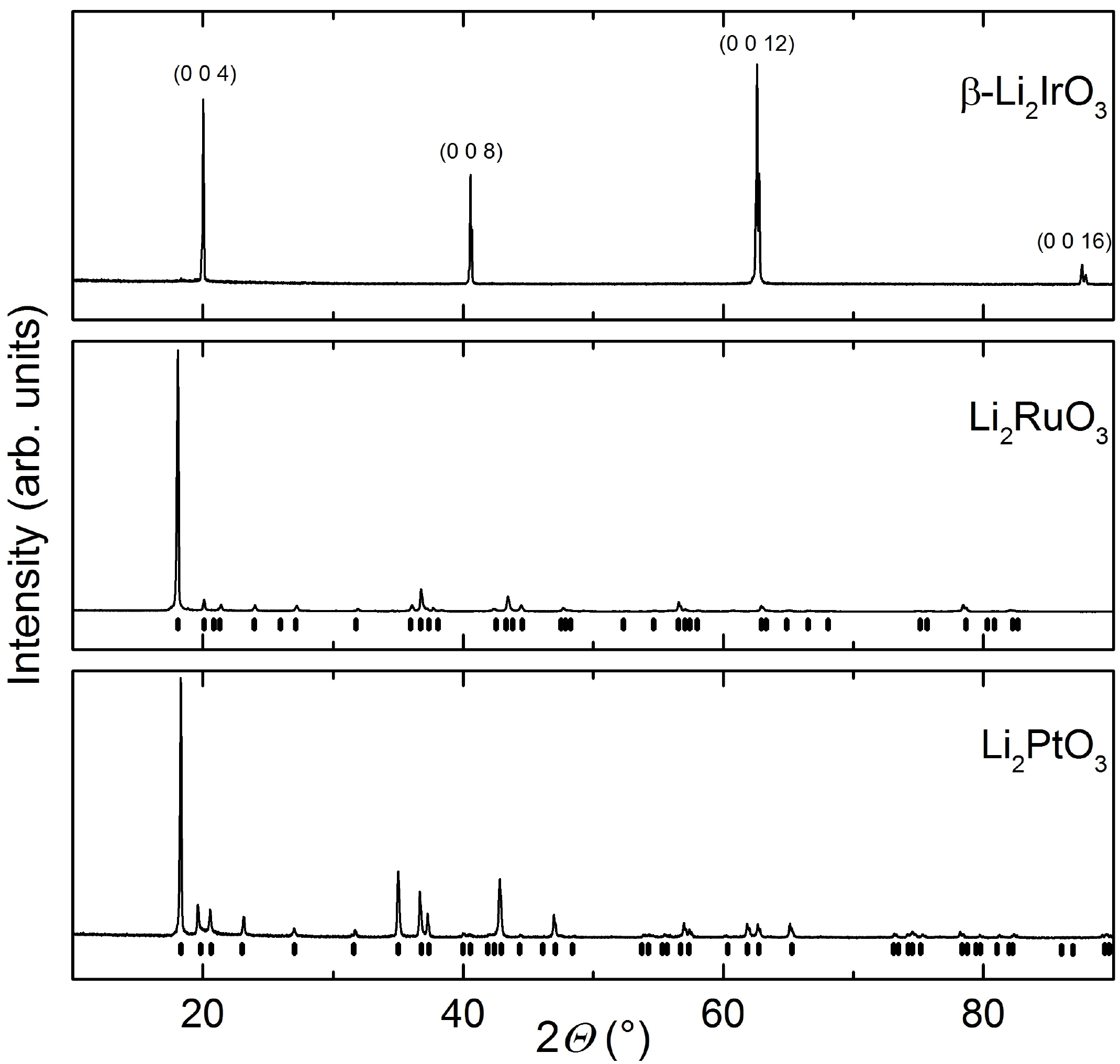}
	\caption{X-ray diffraction patterns of \bli, \lipt and \liru using a Rigaku powder diffractometer. A single crystal of \bli was aligned with the (00l) plane satisfying the Bragg`s law. For \liru single crystals were crushed, while \lipt crystallized in well ordered powder. The samples are phase pure and all peaks can be indexed based on the reported crystal structures (black bars below the diffraction patterns). All samples are synthesized using separated educts (Li, Ir/Ru/Pt).}
	\label{fig:sub5}
\end{figure}
\begin{figure}[hp]
\begin{center}
		\includegraphics[width=0.8 \textwidth]{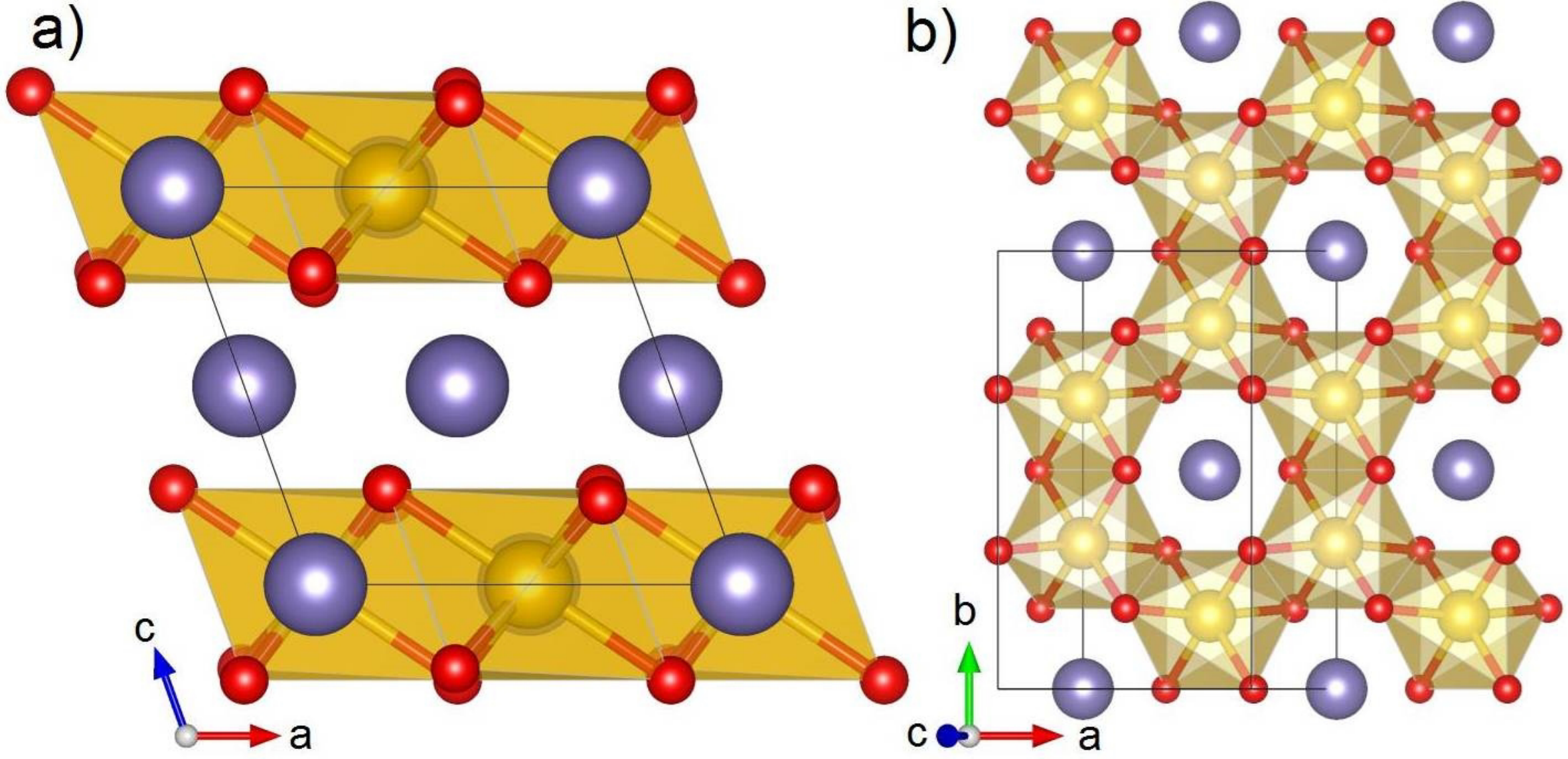}
	\caption[]{Refined crystal structure of $\alpha$-Li$_2$IrO$_3$, showing unit cell as a black outline, Ir as gold spheres, Li as purple and O as red. a) Projection onto the $ac$ plane, showing the alternate
		stacking of Li$_{1/2}$IrO$_3$ and Li layers. b) Basal plane layer
		showing the honeycomb of edge-sharing IrO$_6$ octahedra.}
	\label{fig:struct}
\end{center}
\end{figure}
\begin{figure}[hp]
\begin{center}
\includegraphics[width=0.8 \linewidth]{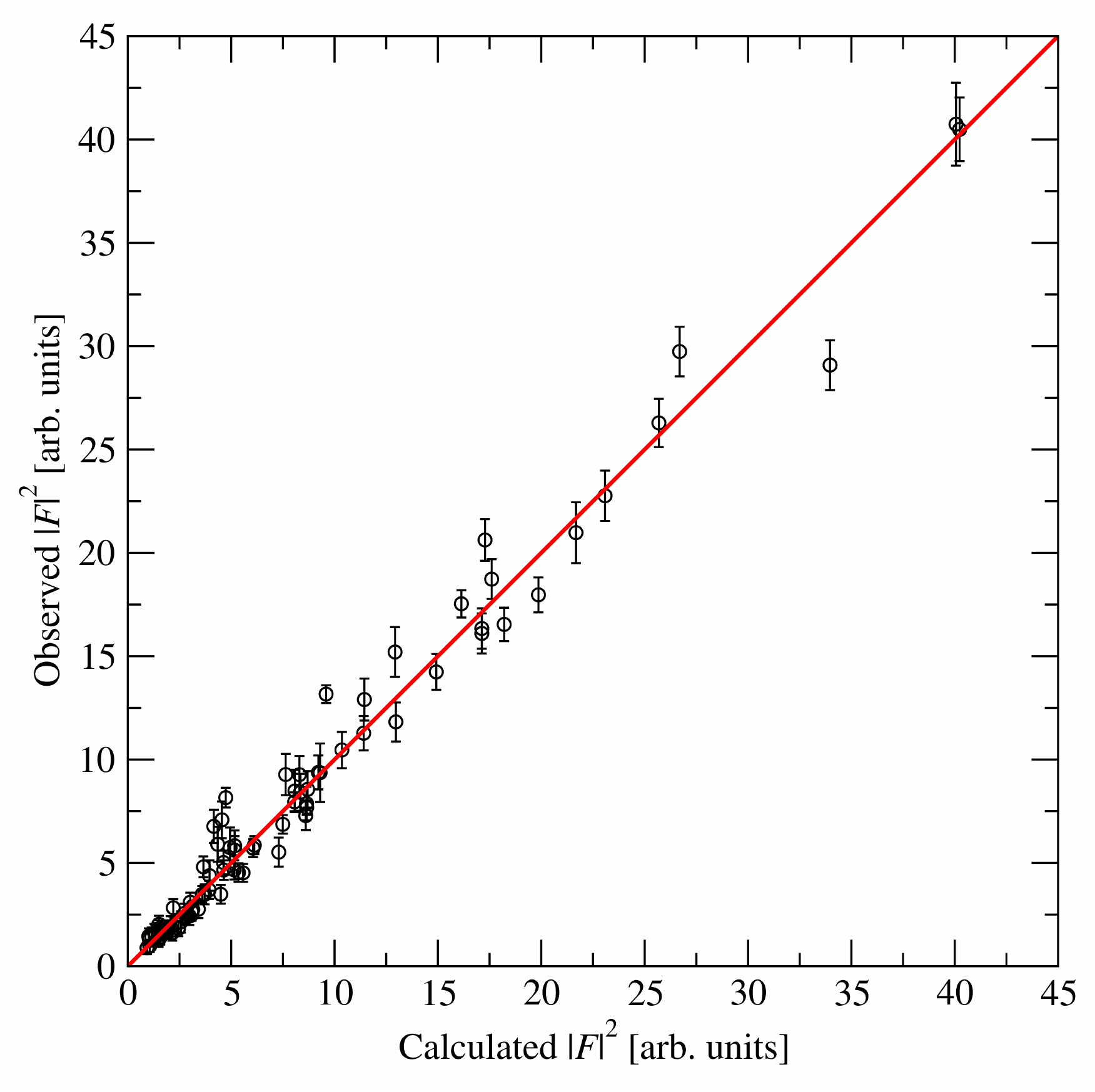}
\caption{\label{fig:comp} Observed vs calculated $|F|^2$ values from Rietveld refinement of the $C2/m$ crystal structure against the single crystal diffraction data. The red line is a guide to the eye.}
\end{center}
\end{figure}

\clearpage
\noindent \textbf{\Large{Supplementary Note}}\\
\\
\textbf{Crystal Structure of \alio}\\
The crystal structure of $\alpha$-Li$_2$IrO$_3$ is shown in Suppl. Fig. 
\ref{fig:struct} and consists of an alternate stacking of honeycombs of
edge-sharing IrO$_6$ octahedra with Li in the center, and
hexagonal Li layers. Adjacent layers are stacked with an in-plane
offset, leading to a monoclinic $C2/m$ space group
\cite{O'Malley08}.
Diffraction data were collected on `Sample 1' (samples are described in the main text, alongside representative data, shown in Fig 4.a-b).
The diffraction pattern could be
fully indexed with a monoclinic unit cell with lattice parameters
(Supplementary Table \ref{struc_tab}) consistent with earlier reports \cite{O'Malley08}, with the
empirical selection rules for the observed Bragg peaks $h + k =
\mathrm{even}$, as expected for a C-centered cell in the $ab$
plane. Importantly, this sample showed no observable diffuse scattering, which would constitute qualitative evidence for stacking faults, nor was there evidence for twinning (discussed below). This allowed for a single phase structural model to be considered without introduction of twin
models or site-mixing approximations for disorder.\\
\\
We performed a full structural refinement of the crystal structure parameters against all integrated
x-ray diffraction intensities. The refinement gave a 
good agreement between model and data as shown in Suppl. Fig. \ref{fig:comp}, with reliability factors $R(F^2)=9.83\%$, 
$wR(F^2)=14.6\%$, $R(F)=5.82\%$ and $\chi^2=3.82$. The refined room 
temperature crystallographic parameters and quantitative details of the 
data collection are given in Supplementary Table \ref{struc_tab}, and the 
corresponding structure is drawn in Suppl. Fig. \ref{fig:struct}. In addition, we 
tested whether or not the crystal structure had lower symmetry than 
$C2/m$ by describing it within the corresponding $P\bar{1}$ 
primitive unit cell. This unit cell preserves all translational symmetry 
of the lattice, but only contains inversion symmetry. In this case, no 
improvement in the fit was achieved, and the same result was found 
within error as for the higher-symmetry, nominal $C2/m$ cell in Supplementary Table \ref{struc_tab}. Further tests to explore Ir disorder through site-mixing 
with the Li ions also showed no improvement in the refinement, leading 
us to the conclusion that Sample 1 has a fully ordered, layered iridium 
honeycomb lattice in all three dimensions. We note that a key difference 
between the solution presented herein and that published by O'Malley
$et~al.$ is the $z$-coordinate of O$2$. In our solution, the oxygen ions 
lie more closely within a common $ab$-plane, as observed in 
Na$_2$IrO$_3$ \cite{Choi12}.

\begin{table}
\begin{center}
\caption{\label{struc_tab}$\alpha$-Li$_2$IrO$_3$ room temperature crystal structure. Standard errors on refined parameters are given in parenthesis. Note that lithium parameters were fixed to literature values \cite{O'Malley08}.}
\vspace{5mm}
\begin{tabular}{c c c c c c}
\toprule
\multicolumn{5}{l}{\textbf{Cell parameters}} \\
\multicolumn{5}{l}{Space group: $C2/m$} \\
\multicolumn{5}{l}{Z = 4} \\
$a,b,c$ ($\mathrm{\AA}$) & 5.1749(2)  & 8.9359(2) & 5.1194(2) &\\
$\alpha,\beta,\gamma$ ($^\circ$)& 90 & 109.827(5) & 90 &\\
\multicolumn{5}{l}{Volume: 222.70(2) $\mathrm{\AA}^3$}\\
\\
\multicolumn{5}{l}{\textbf{Atomic fractional coordinates}} \\
Atom & $x$ & $y$ & $z$ & U$_\mathrm{iso}$($\mathrm{\AA}^2$ ) \\
\hline
Ir1 & 0 & 0.3331(4) & 0 & -\\
Li1 & 0 & 0 & 0 & 0.00633 \\
Li2 & 0 & 0.809 & 0.5 & 0.00633 \\
Li3 & 0 & 0.5 & 0.5 & 0.00633 \\
O1 & 0.25(1) & 0.321(5) & 0.765(8) & 0.03(1) \\
O2 & 0.27(2) & 0 & 0.76(1) & 0.03(1)\\
\\
\multicolumn{5}{l}{\textbf{Ir anisotropic adps ($\mathrm{\AA}^2$) }} \\
U$_{11}$ & 0.019(7) \\
U$_{22}$ & 0.007(2) \\
U$_{33}$ & 0.028(3) \\
U$_{12}$ & - \\
U$_{13}$ & 0.011(3) \\
U$_{23}$ & - \\
\\
\multicolumn{5}{l}{\textbf{Data collection}} \\
\multicolumn{5}{l}{$\#$ measured reflections: 2415} \\
\multicolumn{5}{l}{$\#$ independent reflections: 83} \\
\multicolumn{5}{l}{Data reduction $R_\mathrm{int}$: 5.76\%} \\
\multicolumn{5}{l}{(Criterion for observed reflections: $I > 2.0\sigma(I)$)} \\
\multicolumn{5}{l}{$\#$ observed reflections: 83} \\
\multicolumn{5}{l}{$\#$ fitted parameters: 12} \\
\bottomrule
\end{tabular}
\end{center}
\end{table}
\newpage
\noindent \textbf{Stacking Faults in \alio}\\
As discussed in detail in the case of the iso-structural
Na$_2$IrO$_3$ \cite{Choi12}, as a result of the near hexagonal
symmetry of the honeycomb layers there is only a rather small
energy cost associated with a honeycomb layer being displaced
in-plane by $\pm \textbf{b}/3$. The presence of such occasional
in-plane displacements in the layer stacking sequence is
manifested in the X-ray diffraction pattern by diffuse rods of
scattering along the $l$ direction with the selection rule $k =
3n+1$ or $3n+2$ ($n$ integer) and $h+k$ = even (due to the
$C$-centering). The diffraction patterns of two samples affected by stacking faults 
to different extents are compared in
Fig.\,4b$-$c in the main text and in the intensity profile in
Fig.\,4g in the main text. It is clear that while sample 1 shows very
strong peaks at integer $l$ with negligible diffuse signal
in-between the sharp peaks, by contrast for sample 2 the Bragg
peaks are very weak relative to the strong diffuse scattering,
which extends across a wide range of $l$ values.\\
\\
\textbf{Twinning in \alio}\\
There is a high incidence of twinning in \alio crystals. Most
common are twins with the $ab$ planes rotated relative to one
another by $120^{\circ}$ around the normal direction, $\textbf{c}^*$.
The occurrence of such twins can be understood as follows. In the
nominal structure subsequent honeycomb layers are vertically
stacked with an in-plane offset along the $-\textbf{a}$ direction, see Supplementary
Figure \,\ref{fig:struct}a. Since the honeycomb layers have a
near three-fold rotational symmetry around the vertical axis, there would be
only a small energy cost if the in-plane offset direction were to rotate in the $ab$ plane by $\pm120^{\circ}$, and, if subsequent
layers were then to stack according to this new offset direction, a
new grain would form, rotated around the $\textbf{c}^*$ direction of
the primary grain. We label such a twin C$^{\pm}$, where the
superscript indicates the sign of the $\pm120^{\circ}$ rotation.\\
Another common twin (designated as type A) corresponds to a
rotation by $180^{\circ}$ around the $(101)$ direction, which
swaps the $\textbf{a}$ and $\textbf{c}$ axes. This might occur because the
$a$ and $c$ lattice parameters are very similar in magnitude. A
third type of twin found corresponds to a rotation by
$\pm90^{\circ}$ around the $(\bar{1}01)$ axis, designated as
B$^{\pm}$. We note that for $a=c$ an A-type rotation is equivalent
to two consecutive B-type rotations followed by a $180^{\circ}$
rotation around (010), i.e. ${\rm A}= 2_{y}({\rm B}^{\pm})^2$. For
the given crystal structure the two-fold rotation around $\textbf{b}$
leaves the atomic arrangement invariant as it is a symmetry
operation of the $C2/m$ space group.\\
The above twinning types are summarized in Supplementary Table~\ref{tab:twins},
where the last column gives the transformation matrix between the
reciprocal space coordinates $(h',k',l')$ of the rotated twin and
the corresponding coordinates of the primary (un-rotated) twin,
defined as
\begin{equation}
(hkl)=(h'k'l')\cal{M}. \label{eq:M}
\end{equation}
For simplicity the transformation matrix is given in the case when
the structure has a hexagonal metric, where $a:b:c=1:\sqrt{3}:1$
and $\cos\beta=-1/3$ (the latter equation corresponds to eclipsed
stacking at the third layer). Those conditions are satisfied in
the actual crystal structure to better than $1.4\%$.\\
Fig.\,4d in the main text shows the diffraction pattern for a
multi-twinned crystal. Using the transformation matrices for the
various types of twins we identify the peaks near fractional
positions $(0,2,n+1/3)$ ($n$ integer) as being the nominal
$(1\bar{1}n)$ reflections of a C$^{+}$ twin, whereas peaks near
$(0,2,n-1/3)$ are the nominal $(\bar{1}\bar{1}n)$ reflections of a
C$^{-}$ twin. Similarly, the peaks near $(0,k,\pm1)$, $k$ = odd
are identified as being the nominal $(\pm1,-k,0)$ reflections 
\begin{table}[t]
\caption{\label{tab:twins} Common twinning types in \alio
crystals. Last column gives the transformation matrix ${\cal M}$
for the reciprocal space coordinates as defined in (\ref{eq:M}).}
\vspace{-0.2cm}
\par
\begin{center}
\begin{tabular}{c|c|c}
Type & Orientation & $\cal{M}$\\
\hline
 A & \renewcommand{\arraystretch}{2.2}
\parbox{0.4\linewidth}{\centering rotation by $180^{\circ}$
around ($101$), exchanges $\textbf{a}$ and $\textbf{c}$.} &
\renewcommand{\arraystretch}{2.2}$
\begin{pmatrix} 0 & 0 & 1\\ 0 & -1 & 0\\ 1 & 0 & 0 \end{pmatrix}$\\
\hline
 B$^{\pm}$ & \parbox{0.4\linewidth}{\centering rotation by
$\pm90^{\circ}$ around ($\bar{1}01$), brings the $\pm\textbf{b}^*$
axis along $(101)$} &
\renewcommand{\arraystretch}{1.8}$\begin{pmatrix} 1/2 & \pm 3/2 &
-1/2\\ \mp 1/3 & 0 & \mp 1/3\\ -1/2 & \pm 3/2 &
1/2 \end{pmatrix}$\\
\hline C$^{\pm}$ & \parbox{0.4\linewidth}{\centering rotation by
$\pm120^{\circ}$around ($001$)} &
\renewcommand{\arraystretch}{1.8}$
\begin{pmatrix}   -1/2 & \pm 3/2 & 1/2 \\ \mp 1/2 & -1/2 & \pm 1/6 \\ 0 & 0 & 1 \end{pmatrix}$\\
\end{tabular}
\end{center}
\vspace{-1cm}
\end{table}
\noindent 
of an A twin. All peaks seen in the diffraction pattern can then be
accounted for by an appropriate weighting of a primary grain with
A, C$^+$ and C$^-$ twins, compare Fig.\,4d and h) in the main text.
The presence of B$^\pm$ twins is most easily revealed by the
diffraction pattern in the ($h0l$) plane (not shown) via peaks at
$(2n/3,0,2n/3)$ ($n$ integer), which correspond to nominal
($0,\mp2n,0$) reflections of the rotated grain. The B$^+$ or B$^-$
twins can subsequently be distinguished for example by the
diffraction pattern in the ($\bar{h}kh$) plane (not shown) where
the nominal (001) reflection of the twins will appear at
($-1/2,\pm3/2,1/2$), respectively. We note that all the above
types of twins could also occur in combination, i.e. an A type
twin of the primary grain may have its own C$^{\pm}$ twins and so
on.

\end{document}